\documentclass[english,12pt,a4paper]{article}
\usepackage{babel,amssymb,amsmath,graphicx,hyperref}

\begin{document}

\begin{center}
{\LARGE  Silhouettes of wormholes traversed for radiation}\\~\\
{\large  S.V. Repin$^{1}$, M.A. Bugaev$^2$,  I.D. Novikov$^{1,3,4}$, I.D. Novikov jr.$^1$}
\end{center}
\begin{center}
    \textit{$^1$AstroSpace Center of P.N. Lebedev Physical Institute, \\
              84/32, Profsoyusnaya str., Moscow, 117997, Russia}
\end{center}
\begin{center}
\textit{$^2$Moscow Institute of Physics and Technology,  \\
             9, Institutskiy per., town Dolgoprudnyi, Moscow region, 141700, Russia}
\end{center}
\begin{center}
    \textit{$^3$The Niels Bohr International Academy, The Niels Bohr
             Institute, \\
             Blegdamsvej 17, DK-2100, Copenhagen, Denmark}
\end{center}
\begin{center}
    \textit{$^4$National Research Center Kurchatov Institute,    \\
             1, Akademika Kurchatova pl., Moscow, 123182, Russia}
\end{center}

\bigskip

\begin{abstract}
         The problem of the passage of light through the mouth of a zero-mass wormhole and the possibility of observing
         the objects from another asymptotically flat space-time through the mouth of a wormhole are considered. It is 
         shown that an individual star can have several images and the fact that the image of a flat Lambertian screen has 
         a complex brightness distribution for an observer located on the opposite side of the throat. Images of two such 
         screens visible inside the silhouette of a massless wormhole and the distribution of radiation intensity in their images 
         are constructed.
\end{abstract}


\section{Introduction}
         In papers \cite{Bugaev_2021, Bugaev_2022a, Bugaev_2022b,Novikov_2021b} we started a program of 
systematic comparison of silhouette\footnote{In numerous works on the bending and scattering of light rays by black 
holes and wormholes, the resulting dark formations are called either ``shadows'' 
\cite{EHT_collaboration_2019a, Roy_2020, Mikheeva_2020}, or ``silhouettes'' \cite{Lacroix_2017, Vetsov_2018}. We 
consider the term ``silhouette'' more appropriate and will continue to use it.} features~--- the dark spots arising from 
the observation of radiation interacting with black holes (BHs) and wormholes (WHs). To elucidate and interpret these 
features, it is advisable to start with the simplest models of radiation sources and the simplest WH models. In this approach, 
the features associated with the different nature of these objects, and not with additional effects caused by 
the complexities of more realistic models, are most clearly manifested.

       In this work, we will continue to use this approach. In \cite{Bugaev_2021} we used the Ellis--Bronnikov--Morris--Thorne 
WH model \cite{Ellis_1973, Bronnikov_1973, Morris_1988a, Morris_1988b}, assuming that the WH is filled with an opaque 
substance and considering only the rays moving in our space around the entrance of a wormhole located in the space of 
the observer. In this work, we take into account the light passing through the WH from another space, i.e. we consider 
the WH to be freely traversable for light. See links to previous works in~\cite{Bugaev_2021}. The equations of motion are 
also derived there (see \cite{Zakharov_1999} for a method) and the description of our numerical methods is given.

       The wormhole metric we are considering can be written as:
\begin{equation}
          ds^2 = c^2 dt^2 - dR^2 - \left(R^2 + q^2\right) \left(d\theta^2 + \sin^2\theta\,d\varphi^2\right)\,.
          \label{MT_metric}
\end{equation}
Here $c$ is the speed of light. In what follows, we set $c = 1$.  We further set $c = 1$.

\section{Formulation of the problem}

       First of all, we will look at the images and silhouettes (shadows) that are created (formed) by the rays that have passed 
through the wormhole.

\begin{figure}[htb]
  \centerline{
  \includegraphics[width=12cm]{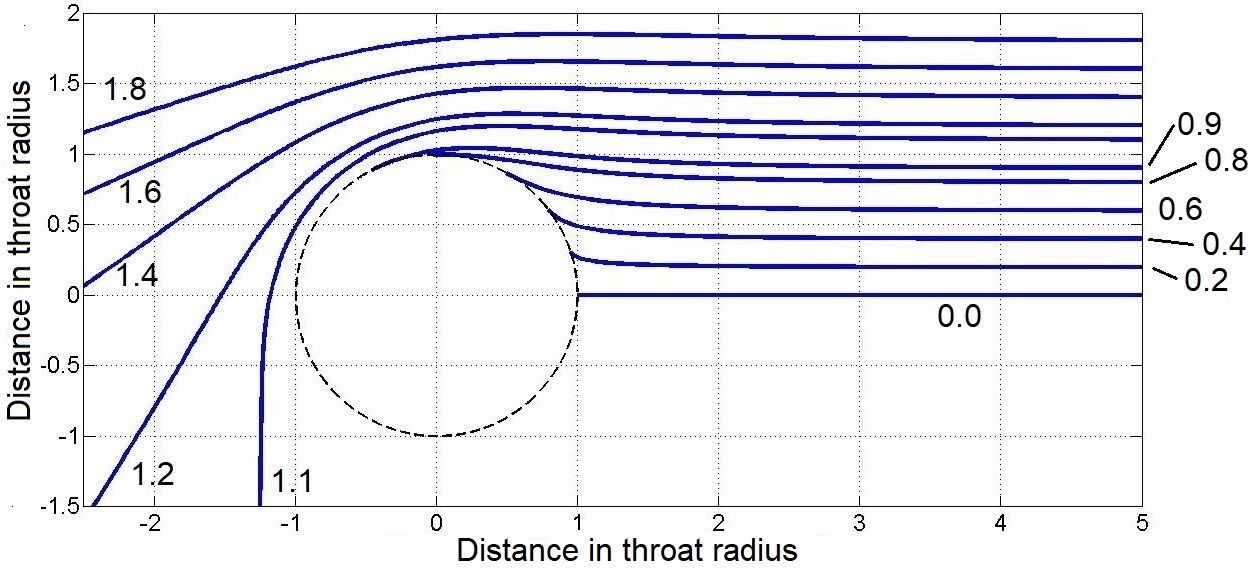}
             }
 \caption{Section of observer space (spacetime-1) with $\theta = 90^\circ$ (equator), \mbox{$0 < R < \infty$}, 
                 $0 < \varphi < 2\pi$. The trajectories of null geodesics (rays of light) in this space are given. These trajectories 
                 extend from the observer to the right at infinity $R \to \infty$, $\varphi = 0$ to the wormhole entrance area. 
                 Trajectories have the impact parameters $b = 0$, 0.2, 0.4, 0.6, 0.8, 0.9, 1.1, 1.2, 1.4, 1.6 and 1.8. 
                 The circumference is the throat of a wormhole. The system coordinates correspond to 
                 the formula~(\ref{MT_metric}). Distances along the vertical and horizontal axes are plotted in units of the size 
                 of the throat radius.}
  \label{WHMT_trajectories}
\end{figure}

\begin{figure}[!htb]
  \centerline{
  \includegraphics[width=6.8cm]{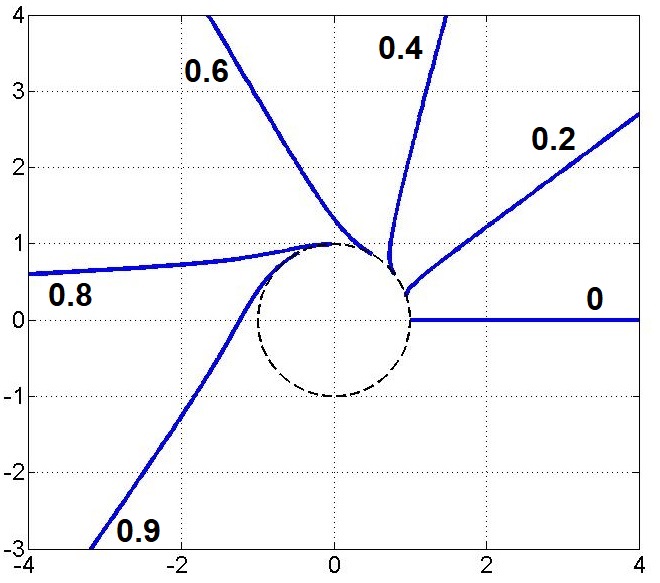}
             }
  \caption{Spacetime-2.  Continuation of trajectories with impact parameters: 0, 0.2, 0.4, 0.6, 0.8 and 0.9 
                in space-2 after crossing the throat. Distances along the coordinate axes are measured in units of 
                the throat radius.}
  \label{WH_trajectories_space2}
\end{figure}

\begin{figure}[!hbt]
  \centerline{
  \includegraphics[width=12cm]{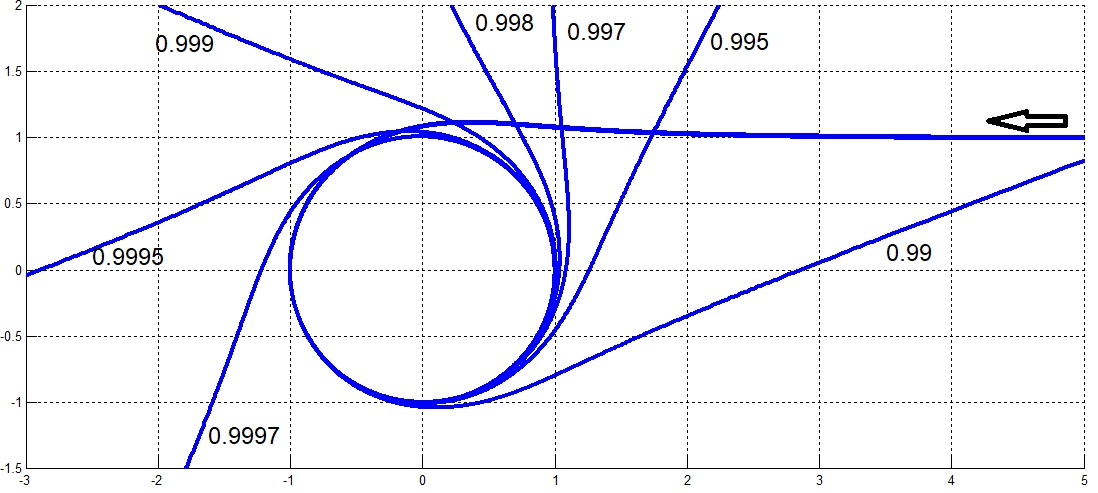}
             }
  \caption{Trajectories of zero geodesics in space-2, strongly curving around the throat. The impact 
                 parameter is indicated next to the trajectory.The horizontal trajectory next to the arrow is 
                 the incoming trajectories in space-1 that merged at a given resolution for all the trajectories 
                 shown in the figure.}
  \label{WHMT_close_1}
\end{figure}

        We emphasize that in the metric of a spherical wormhole, as well as in the metric of a spherical black hole, 
the spatial trajectories of light rays are always flat. Therefore, it will suffice for us to consider only one plane section of 
the spatial metric, for example, the equatorial one.  For an observer who is far enough away from the entrance of 
the wormhole, the rays coming to him from the entrance and its surroundings are almost parallel. Note that 
the trajectories of null geodesics in the metric (\ref{MT_metric}) for $|R| \gg 1$ quickly become almost straight lines. 
Figure~\ref{WHMT_trajectories} shows the outer space of the observer in the equatorial section $\theta = 90^\circ$ (in 
the coordinates of the formula (\ref{MT_metric})). The circle is the throat of the wormhole, where $R = 0$. The observer 
is at $\varphi = 0$, $R \to \infty$.  We will call the space $0 < R < +\infty$ as space-1. A parallel beam of null geodesics 
leaves the wormhole region towards the observer. Let's call it the sheaf $A$.

         Figure~\ref{WHMT_trajectories} shows the trajectories of null geodesics with different impact parameter $b$ with 
respect to the observer. The parameter $b$ is measured in units of the throat radius. Zero geodesics with $b > 1$ are 
the rays in the space-1 of the observer. Trajectories with $b < 1$ reach the WH throat and go to another space. We will 
call it space-2, where $R$ changes from $R = 0$ to $R \to -\infty$.

         Figure~\ref{WHMT_trajectories} shows the trajectories with impact parameter~$b$ above the horizontal line 
$b = 0$, i.e. moving counterclockwise. Of course, a similar picture should be repeated for the impact parameter~$b$ 
below the line $b = 0$, i.e. moving clockwise.

        Figure~\ref{WH_trajectories_space2} shows space-2.  Shown here are null geodesics that arrive 
from space-1 crossing the throat. The corresponding impact parameters $b < 1$ are indicated next to the trajectories. 
For values $b < 1$, but very close to 1 (i.e. $(1 - b) \ll 1$) the trajectories turn many times around the $\varphi$ 
coordinate before going to $R \to -\infty$. This is shown in fig.~\ref{WHMT_close_1}.

         Trajectories with $b < 1$ are the object of our attention in this article. They are used to build the images and 
silhouettes that are visible to the observer through the wormhole.

\section{Images of individual stars seen through a wormhole}

        Let us now proceed to our main task - to construct the images of individual objects and silhouettes visible to 
the observer through the wormhole. We start with the image of individual stars.

\begin{figure}[hbt]
  \centerline{
  \includegraphics[width=6.7cm]{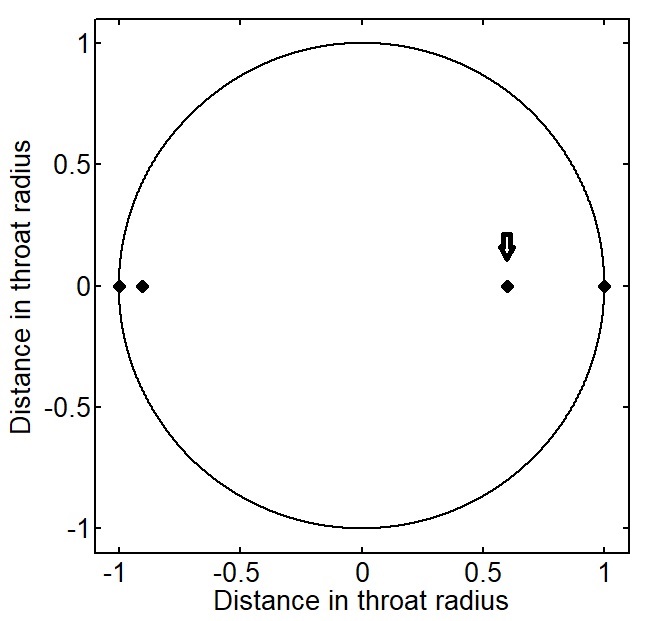}
  \includegraphics[width=7.7cm]{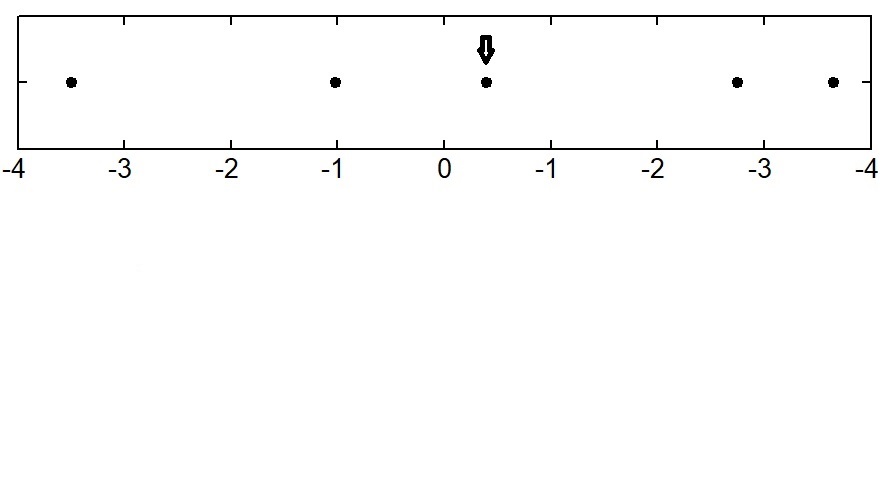}
             }
  \caption{Multiple images of the same star in space-2. The image marked with an arrow can be 
                 conventionally called the ``main'' image.  In the right panel, the same images are shown 
                 with the horizontal coordinate $\log(1-|b|)$.}
  \label{Star_through_WH_1}
\end{figure}

        Let there be a distant point source (star) in space-2 located in our $R - \theta$ section at the coordinate~$\theta$. 
How will it appear to the observer?  Rays of light from a distant star (zero geodesic) fall on the entrance of the wormhole 
in space-2 in a parallel beam. Let's call it the sheaf $B$. A ray of light, the asymptotics of which for $R \to -\infty$ 
corresponds to the coordinate~$\theta$,  will reach the observer. Its impact parameter~$b_1$ will correspond to 
the position of the star image for the observer. But the image created will not be a single one.
A ray from beam $B$ going near the throat and making a full turn near it can get into the beam $A$ and come to 
the observer. This is how the second image of the star $b_2$ appears closer to the throat circumference, and so on. 
There will be an infinite number of images, getting closer and closer to the circle corresponding to the throat. The rays from 
beam $B$ can also make full turns in the opposite direction. Such images will also be observed near the throat circumference, 
but near its opposite side. Multiple images of a single star are shown in Fig.~\ref{Star_through_WH_1} on the left. 
The same images are shown on the right, but the value $\log(1-|b|)$ is used as the horizontal coordinate. This is done 
in order to distinguish individual merging images near the edge of the throat.

\begin{figure}[hbt]
  \centerline{
  \includegraphics[width=6.7cm]{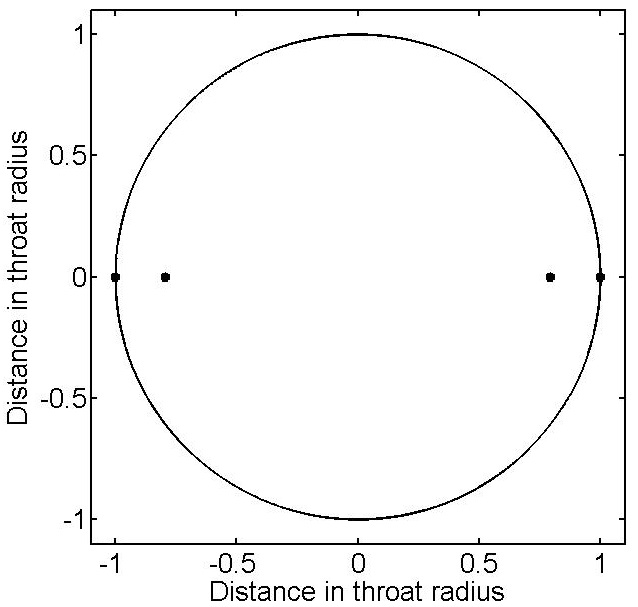}
  \includegraphics[width=7.7cm]{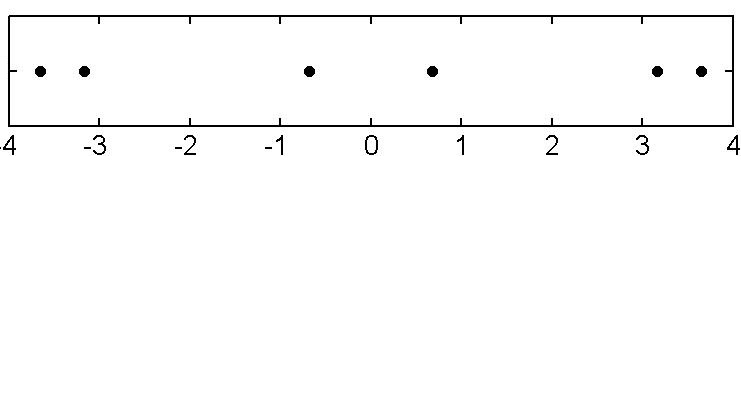}
             }
  \caption{Multiple images of the same star located in space-2 and arranged symmetrically in the observer's 
                 field of view. For the same images in the right panel, the $\log(1-|b|)$ coordinate is used.}
  \label{Star_through_WH_2}
\end{figure}

\begin{figure}[!hbt]
  \centerline{
  \includegraphics[width=6.7cm]{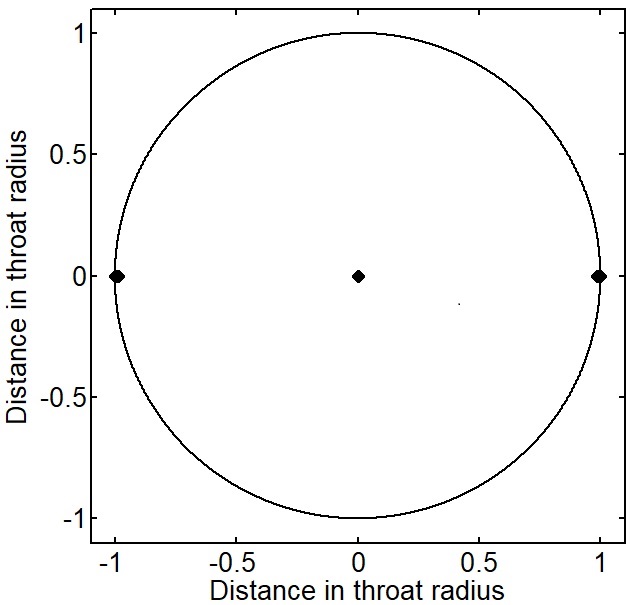}
  \includegraphics[width=7.7cm]{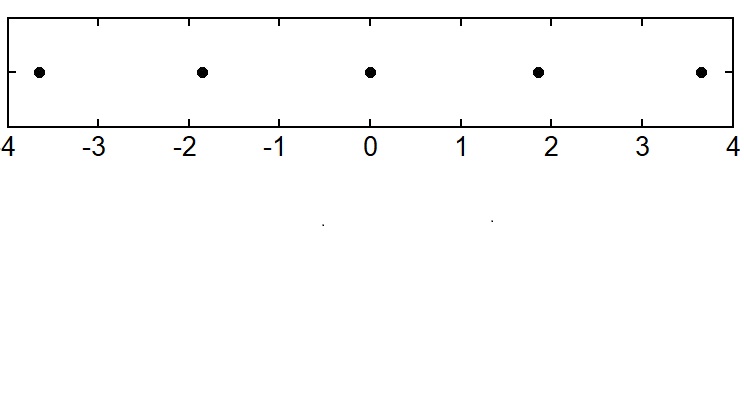}
             }
  \caption{Multiple images of the same star located in space-2 and arranged symmetrically in the observer's 
                 field of view. The ``main'' image is in the center of the field of view. For the same images in 
                 the right panel, the $\log(1-|b|)$ coordinate is used.}
  \label{Star_through_WH_3}
\end{figure}

      The nature of the image of the star itself and its repetition depend, of course, on its position in the sky of space-2.

      We will assume that the meridian of the section $\varphi = 90^\circ$, along which we consider the different positions 
of the stars, corresponds to a horizontal line in the field of view of the wormhole image observer. Then the pattern with 
repeated images is shown schematically in Fig.~\ref{Star_through_WH_2}. In the region of the pole, the first image of 
the star is absent. The rays from this star do not follow a straight path directly to the observer. At the same time, 
the outermost points in the field 
of view in the left panel consist of two merged images that cannot be resolved at this scale. On the right panel with 
the horizontal coordinate $\lg(1-|b|)$ they are quite clearly visible.

      Fig.~\ref{Star_through_WH_3} shows another symmetrical situation, when the star from space-2 is visible in 
the center of the field of view.

\section{Inner silhouettes (shadows) of a wormhole}

\begin{figure}[!htb]
  \centerline{
  \includegraphics[width=10cm]{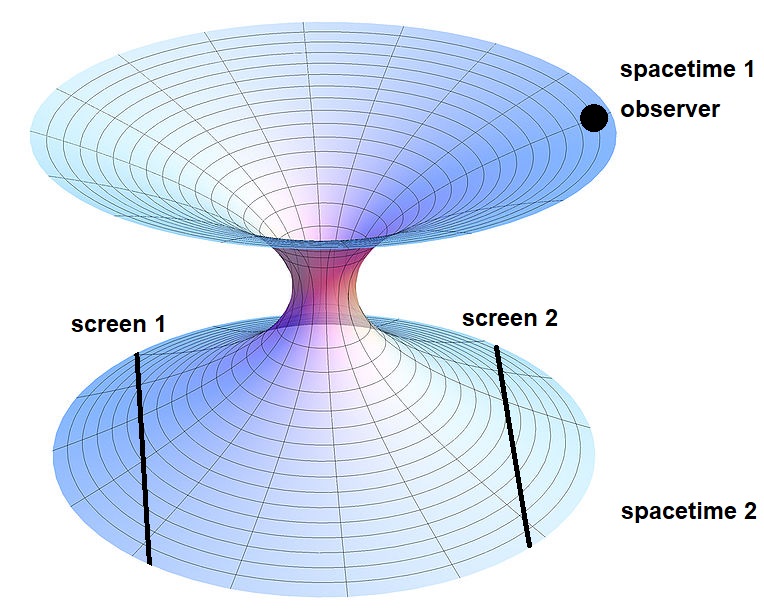}
             }
  \caption{Layout of the observer in space-1 and extended Lambertian screens in space-2.}
  \label{WH_image_in_1}
\end{figure}

      Let us now consider the images created for the observer in our space-1 by extended luminous screens located in 
space-2. As in the papers~\cite{Bugaev_2021}--\cite{Bugaev_2022b} we will assume that the screen glow intensity is 
the same in all directions (Lambertian source).  First of all, consider a screen located in space-2 in the direction of the pole 
$\varphi = 180^\circ$ perpendicular to the radial direction and located far enough $R \to - \infty$. Fig.~\ref{WH_image_in_1}
schematically shows the position of the observer in space-1 and two screens located in space-2.  It is clear from the Figure 
that the light beam coming from the observer exactly to the center of the wormhole (i.e. with zero impact parameter) arrives 
at screen-2 perpendicular to it and therefore does not hit screen-1. And in order to hit the screen-1, the light ray must move 
with a non-zero impact parameter $b$ and, therefore, turn in space-2 (near the throat) by an angle greater than $90^\circ$.
Looking again at Fig.~\ref{WH_trajectories_space2} one can see that the impact parameter of such a beam is approximately 
$b \approx 0.5$.

\begin{figure}[!htb]
  \centerline{
  \includegraphics[width=8cm]{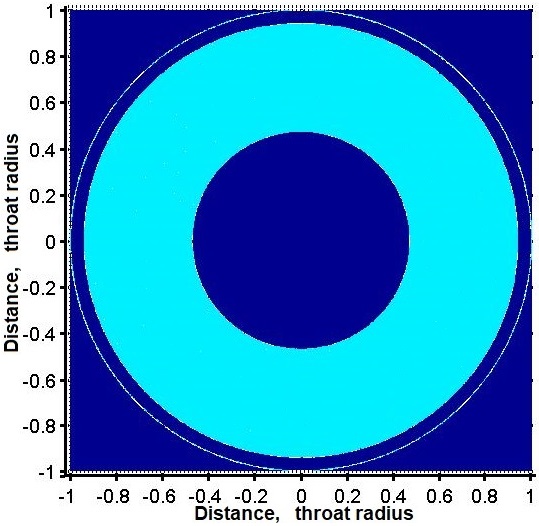}
  \includegraphics[width=9cm]{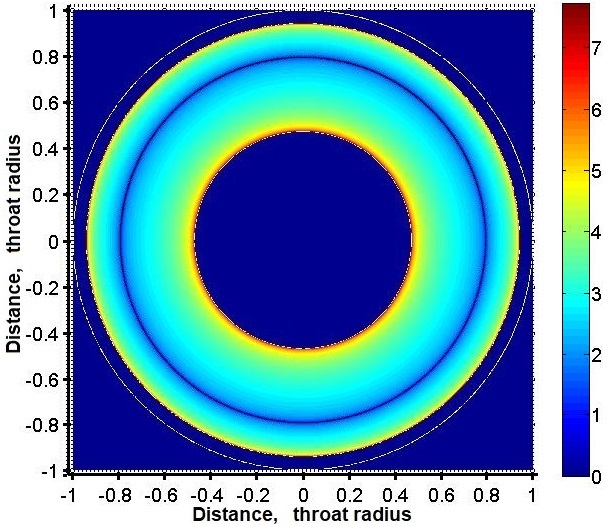}
             }
  \caption{The image of a flat Lambertian screen-1, which the observer sees through the mouth of a wormhole.
                The silhouette of this screen is shown on the left, and the image emission intensity distribution on 
                a logarithmic scale is shown on the right.}
  \label{WH_image_in_2}
\end{figure}

\begin{figure}[!htb]
  \centerline{
  \includegraphics[width=12cm]{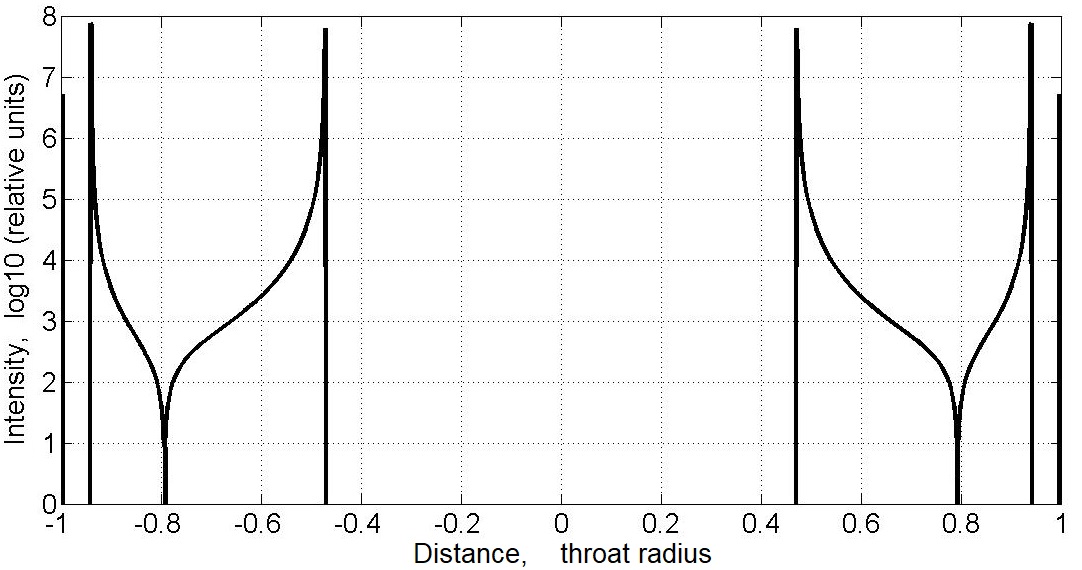}
             }
  \caption{Radiation intensity distribution along the diameter of the wormhole throat for the first screen.}
  \label{WH_distribution_in_2}
\end{figure}

        First of all, we will consider how the image of screen-1 looks like from the point of view of an observer located in 
space-1. We consider the screen dimensions to be infinitely large The methods of constructing the images are similar to those described in the paper~\cite{Bugaev_2021}. The image seen by an observer in space-1 is shown in Fig.~\ref{WH_image_in_2}.
The left panel shows the shape of the image as seen by the observer, while the right panel shows the distribution of 
radiation intensity in the same image, shown on a logarithmic scale. The inner radius of the light region, as mentioned 
above, corresponds to a rotation of the ray in space-2 by the angle of ~$90^\circ$, and the outer boundary corresponds 
to a rotation by $270^\circ$. The exact values of the impact parameters for these values of the rotation angle can only 
be obtained by numerical integration of the zero geodesic equations. These values are:
\begin{equation}
      b_{90} = 0.4704q, \qquad\qquad  b_{270} = 0.94008q\,.
\end{equation}
The beam, however, can reach the screen in a more complicated way: it can make several revolutions near the throat of 
the wormhole and only then hit the screen. So, when turning the ray through the angles from $450^\circ$ to $630^\circ$, 
the ray will reach the screen after making one full turn, and when turning through the angles from $810^\circ$ to 
$990^\circ$ --- two full turns, etc. These rays should pass through the mouth of the wormhole with impact parameters 
very close to unity and form an infinite number of thin rings. For the first such ring, the impact parameters of the inner 
and outer boundaries refer respectively to:
\begin{equation}
      b_{450} = 0.9969445q, \qquad\qquad  b_{630} = 0.9998713q, 
                                              \qquad\qquad  b_{630} - b_{450} \approx 0.003q\,.
\end{equation}
The accuracy of these values is four to five units of the last decimal place. For the second and subsequent rings, 
the integration with extremely high accuracy is required to determine these parameters, since the system of equations 
becomes too stiff for standard numerical methods.

        The intensity distribution along the image diameter is shown as a graph in Fig.~\ref{WH_distribution_in_2}. The local 
intensity minimum falls at the value of the impact parameter $b = 0.793$, which corresponds to the beam rotation 
by~$180^\circ$. The appearance of this minimum can also be easily understood from general physical considerations. 
The point is that the rays emitted from an observer with an impact parameter $b = 0.793$ hit the same point of screen-1 in 
space-2 and, consequently, the observer sees this point when registering any ray with an impact parameter $ b = $0.793.

        Consider now the screen-2, in Fig.~\ref{WH_image_in_1}, located in space-2 at $\varphi = 0^\circ$, $R \to - \infty$. 
In this case, the beam coming with a zero impact parameter hits the center of the screen and, therefore, the center of 
the screen should be visible bright. Moreover, a ray that hits the screen-2 cannot hit the screen-1 and vice versa. 
Conventionally, we can say that the screen-2 image should be ``complementary'' to the screen-1 image shown in 
Fig.~\ref{WH_image_in_2}.

\begin{figure}[!htb]
  \centerline{
  \includegraphics[width=8cm]{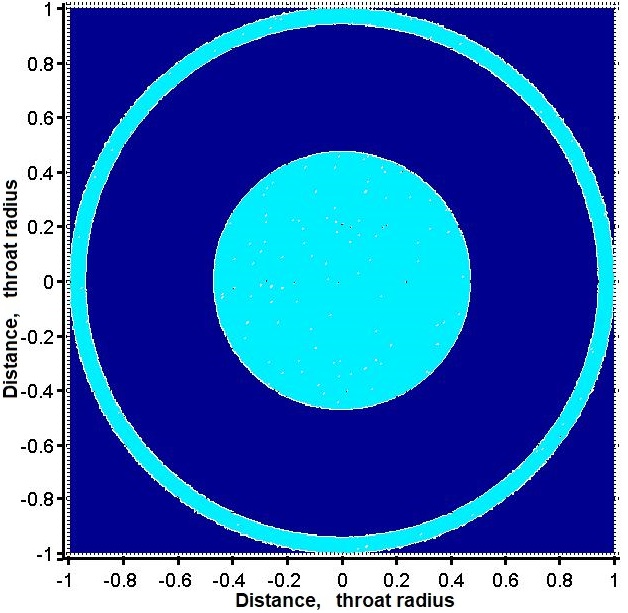}
  \includegraphics[width=9.0cm]{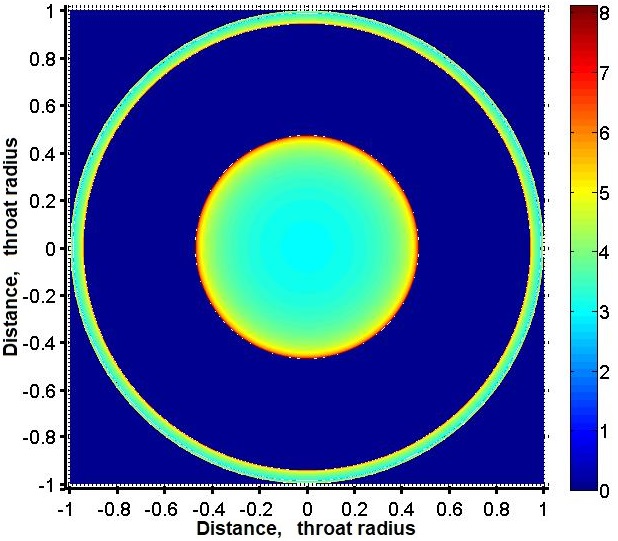}
             }
  \caption{The  image of a flat Lambert screen-2, which the observer sees through the mouth of a wormhole. 
                The silhouette of this screen is shown on the left, and the image emission intensity distribution on 
                a logarithmic scale is shown on the right.}
  \label{WH_image_in_3}
\end{figure}

\begin{figure}[!htb]
  \centerline{
  \includegraphics[width=10cm]{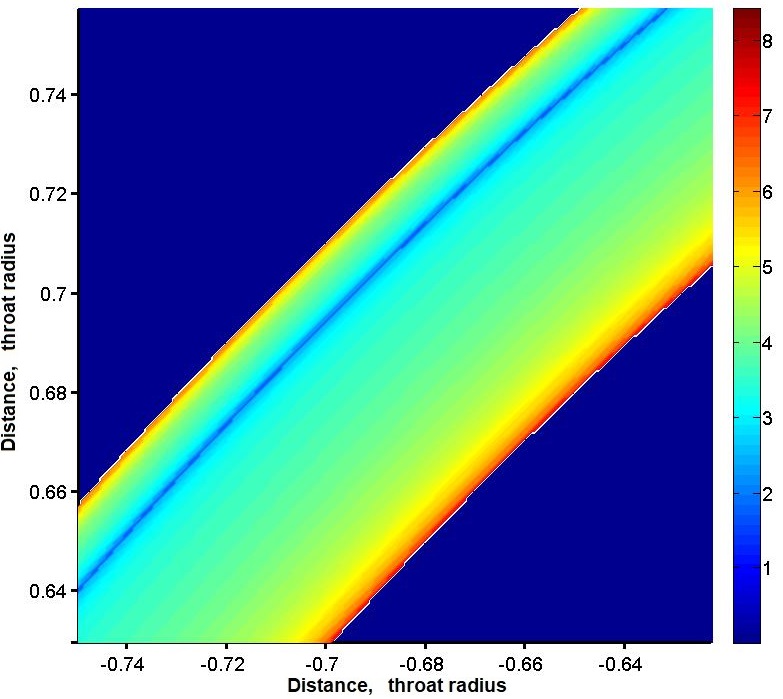}
             }
  \caption{Details of the radiation intensity distribution in the first ring, i.e. in a ring formed by quanta that 
                 have made one complete revolution in the throat of a wormhole.}
  \label{WH_image_in_4}
\end{figure}

\begin{figure}[!htb]
  \centerline{
  \includegraphics[width=12cm]{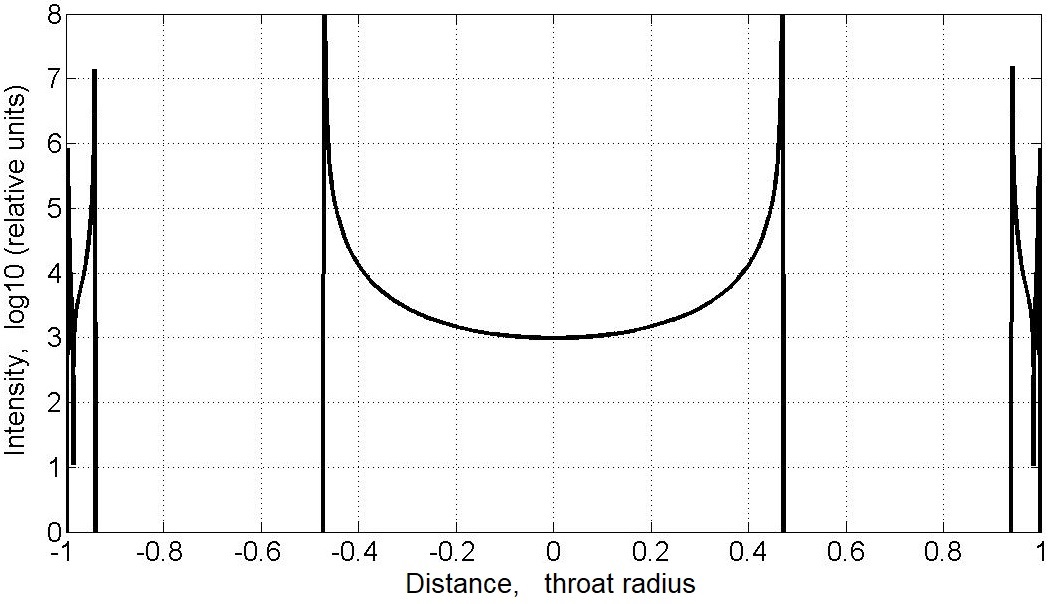}
             }
  \caption{Radiation intensity distribution along the diameter of the wormhole throat for the second screen.}
  \label{WH_distribution_in_3}
\end{figure}

      The image of screen-2 for an observer in space-1 is shown in Fig.~\ref{WH_image_in_3}. Similarly to 
Fig.~\ref{WH_image_in_2}, the left panel shows only the silhouette of the image that the observer sees, while the right 
panel shows the distribution of the radiation intensity in the same image on a logarithmic scale. As expected, the center 
of the screen is bright, but the distribution of brightness differs from Fig.~\ref{WH_image_in_2}. In the center of 
the screen, the intensity has a minimum and increases towards the edge of the bright disk.  In addition, we see a bright 
ring formed by the rays that have made a complete revolution in the mouth of a wormhole. In this case, the width of 
the ring is such that we can plot the radiation intensity distribution inside it along the radial coordinate. On an enlarged 
scale, the intensity distribution in the ring is shown in Fig.~\ref{WH_image_in_4}, where one can see its details. For 
example, the intensity minimum is clearly visible, and this minimum is located asymmetrically and strongly shifted to 
the outer edge of the ring. The radiation intensity increases towards the edges of the ring. As in the previous case, 
the minimum arises due to the fact that all rays with this impact parameter, having made a revolution, fall into the same 
point of the screen. The value of the impact parameter for the intensity minimum in the ring is:
\begin{equation}
      b = 0.985q.
\end{equation}
The graph of radiation intensity distribution along the equatorial section of the wormhole silhouette is shown in 
Fig.~\ref{WH_distribution_in_3}.

        Of course, an observer in space-1 will see both the rays that have passed in his space around the entrance to 
the wormhole, as well as the rays that have passed through the wormhole. We considered the first category of rays 
in~\cite{Bugaev_2021, Bugaev_2022a, Bugaev_2022b}. The complete picture for the observer is formed by merging 
the depictions considered in~\cite{Bugaev_2021, Bugaev_2022a, Bugaev_2022b} and considered in this paper.

\section{Conclusion}

       In the considered model of the silhouette of a traversable wormhole, there are characteristic image details that can 
be used to identify these objects in interferometric observations \cite{Mikheeva_2020}. Thus, in contrast to the black 
holes, fairly complex ring structures with varying image brightness can be observed inside the silhouette of a wormhole.  
Wormholes themselves, like supermassive black holes, may possibly exist in the centers of galaxies \cite{Kardashev_2020}.
In order to search for and observe such objects, one can use the catalog of supermassive black holes~\cite{Mikheeva_2019}, 
in which one can find a sufficient number of candidates for observations using the ground-based interferometers.

         Even wider opportunities for observing the wormholes are provided by a ground-space interferometer, in which one 
of the receiving antennas is located on a satellite located in space at a distance from units to hundreds of Earth radii. One 
of the space objects for installing such an antenna could be the Moon. Another possibility is provided by the Russian project 
Millimetron \cite{Kardashev_2006, Novikov_2021a} with a ten-meter cooled antenna. The resolution of the Millimetron in 
the interferometer mode can reach tens of nano-arcseconds. Such a high resolution is enough to study in detail 
the silhouettes of black holes and, possibly, wormholes in a large number of galaxies. The discovery of such objects will 
undoubtedly give a new impetus to the development of Astrophysics.

\section{Acknowledgments}

      S.R. expresses his gratitude to O.N. Sumenkova, R.E. Beresneva and O.A. Kosareva for the opportunity to successfully 
work on this problem.

\end{document}